\documentclass{article}

\usepackage[left=1.5in,right=1.5in,top=1.5in, bottom=1.5in]{geometry}
\usepackage{graphicx}
\usepackage{epstopdf,epsfig}
\usepackage{newtxtext}
\usepackage{newtxmath}
\usepackage{natbib}
\usepackage{hyperref}
\hypersetup{                            
    colorlinks = true,
    urlcolor   = blue,
    citecolor  = black,
}

\title{Dispersion of a fluid plume during radial injection in an aquifer}

\author{Benjamin W.A. Hyatt$^1$ \and Yuri Leonenko$^1$}
\date{%
    $^1$Department of Earth and Environmental Sciences, University of Waterloo\\
    January 22, 2021
}

\begin{document}
\maketitle

\begin{abstract} 
This study outlines a model for injected fluid flow in a vertically confined porous aquifer with mechanical dispersion. Existing studies have investigated the behaviour and geometry of immiscible fluid flow in this setting, where the injected fluid displaces the resident fluid, forming a sharp interface between the two. The present study extends analytical solutions to include mechanical dispersion of the interface. The solutions are inverted to solve for time as a function of position $(r,z)$, giving each position in the aquifer an intersection time corresponding to the moment the travelling interface intersects a point of interest. The set of $\{r_{o},z_{o}\}$ positions which share an intersection time are treated as dummy variables that represent an “effective surface” and are integrated over to solve for the velocity field within the aquifer. Using this velocity field, the concentration profile resulting from mechanical dispersion can be found analytically. It is shown that the concentration of the injected fluid smoothly decays around the position of the interface from immiscible solutions, allowing for the injected fluid to be present in detectable quantities beyond the extent of these interfaces. This concentration spread should be considered in defining outer boundaries on fluids in injection well applications such as carbon capture and storage or groundwater applications. 
\end{abstract}

\section{Introduction}
\label{sec:intro}

The flow of fluids injected into porous media has been a research area of interest for decades, having wide ranging applications including enhanced oil recovery \citep{BuckleyLeverett42}, assessing drinking water quality \citep{PrommerStuyfzand05} and carbon capture and storage (CCS) \citep{GibbinsChalmers08}. This has inspired the development of many analytical solutions and numerical simulations modelling the behaviour, geometry and evolution of fluids being injected into the resident brine of aquifers under a variety of circumstances. A common feature throughout many of these studies is the assumption that the injected fluid and resident fluid are immiscible; that is, the injected fluid does not dissolve in the resident fluid and forms a sharp interface separating regions of 100\% injected fluid saturation and 100\% brine saturation. Another assumption that can allow for simple approximate solutions for the fluids’ evolution in the aquifer is vertical equilibrium, where the flow velocity in the vertical direction is assumed to be negligible, and the velocity field in the aquifer is strictly radial (in the direction outward from the injection site) \citep{Nordbottenetal05, NordbottenCelia06, Juanesetal10}.

With simplifying assumptions and boundary conditions, analytical solutions for the shape or thickness of this sharp interface as a function of radial distance and time can be obtained. This allows for the furthest radial extent of the interface to be determined after injecting for a certain period. However, sizable traces of the injected fluid may appear beyond this boundary due to mechanical dispersion, a mass transfer mechanism where velocity changes on the scale of the medium’s pores cause the fluid to spread out \citep{Fetter01}: this will cause the concentration of the injected fluid to smoothly decay around the region where this sharp boundary would be. This phenomenon has been examined in studies that assume a radial flow of injected fluid with no variation in the vertical direction \citep{HoopesHarleman67, HsiehYeh14, TangBabu79}. This purely radial flow, however, can be interpreted as having an underlying cylindrical, immiscible interface radiating outward from the injection well, which itself is a special case of a general interface geometry subject to vertical equilibrium. 

The present study proposes a mathematical procedure which allows including the mechanical dispersion process into consideration and applying it to generic interface geometries, extending beyond the cylindrical case. This allows for quantifying a more precise upper boundary cutoff for the distances at which traces of the injected fluid may appear in the aquifer: for example, finding where the relative concentration of the injected fluid drops to some acceptable levels, such as 1\%, 0.5\%, etc. Similarly, a lower boundary distance value for a desired relative concentration cutoff (e.g., 99\%) can be defined to quantify regions in the aquifer that are effectively saturated with the injected fluid. The presence of injected contaminants is of great interest in injection well engineering applications \citep{Ahmadetal10, JankovicFiori10, CombaBraun12, Cahilletal14}. As such, computing these boundaries between regions saturated with groundwater, injection fluid, or a transition zone in between will be of practical in determining injection efficiency and safety parameters. 

For the purposes of this investigation, the resident and injected fluids are assumed to have velocity fields which obey Darcy’s law and are incompressible with constant densities and viscosities. The vertical pressure throughout the aquifer is assumed to be hydrostatic, and the aquifer is assumed to be isothermal and axisymmetric, with a finite height due to impermeable caprock layers at the top and bottom and infinite laterally. Furthermore, the aquifer is assumed to have constant permeability, porosity, and dispersivity scale.

\section{Methodology}\label{sec:method}

The mathematical expression for the incompressibility condition of a fluid states that the divergence of its velocity field is zero \citep{White11}:

\begin{equation}
  \nabla\cdot\boldsymbol{v} = 0
\end{equation}

Assuming an axisymmetric flow with vertical equilibrium, the velocity field in cylindrical coordinates will only have a component in the radial direction, outward from the injection well. To satisfy the incompressibility condition, this radial velocity field must be of the form:

\begin{equation}
  \boldsymbol{v}\propto\frac{1}{r}\hat{\boldsymbol{r}}
\end{equation}

An expression for the velocity field can be obtained from the volumetric flow rate of the fluid $Q$ [L$^{3}$/T] evaluated through some characteristic surface (elaborated on below), which is given by the total flux of the velocity field through that surface \citep{White11}:

\begin{equation}
  Q = \phi\iint\boldsymbol{v}\cdot\boldsymbol{dA} 
\label{Qflux}
\end{equation}

Here, $Q$ is assumed to be equal to a constant volumetric injection rate at the site of the well. The dimensionless factor $\phi$ is the porosity of the aquifer the fluid is flowing through, which is the fraction of volume in the medium that is open and unoccupied of material. This porosity factor is introduced to relate the flow velocity with the volumetric flux from Darcy’s law \citep{HaitjemaAnderson15}. For the purposes of this investigation, the porosity is assumed to be constant throughout the aquifer. 

The nature of the integrated over “characteristic surface” mentioned above is at the heart of this analysis. If the injected and resident fluids are imagined to be immiscible, this surface represents the travelling interface between them the moment it intersects a point of interest $(r,z)$ in the aquifer. Note that this surface should depend only on the intersection position and is independent of time or where the “true” interface is located. Also note that for any point $(r,z)$ there is a corresponding set of points $\{r_{o},z_{o}\}$ which represent the solution for the interface at the time it crosses a point $(r,z)$. These $r_{o},z_{o}$ points represent an effective surface for a point $(r,z)$ and are treated as dummy variables which are used to evaluate the flux integral in equation \ref{Qflux}. 

The velocity field, however, is interpreted to be a function of the specific position $(r,z)$, and not these dummy variables representing an effective surface; as such, it will be factored out of the integral. Additionally, as the velocity field is assumed to be strictly radial, the area element dotted with it can be written as:

\begin{equation}
  \boldsymbol{v}\cdot\boldsymbol{dA} = v(r_{o}d\theta dz_{o})
\label{vDA}
\end{equation}

Where the cylindrical element $r_{o}d\theta dz_{o}$ is the projection of the $\boldsymbol{dA}$ surface element using the dummy variables mentioned above. Consider the example of a cylindrical, immiscible vertical interface travelling through a confined aquifer of height $H$. In this case, $r_{o} = r$ is independent of the vertical $z_{o}$, and evaluating the flux integral in \ref{Qflux} yeilds:

\begin{equation}
  Q = 2\pi\phi rv \int_{0}^{H}dz_{o} = 2\pi\phi rvH 
\end{equation}

Solving for the velocity field above gives $v=Q/2\pi\phi H$, which satisfies the incompressibility condition and is the same equation as the velocity given by \citet{TangBabu79} and \citet{HoopesHarleman67}.

In general, the interface formed by the injected fluid may not have a flat, cylindrical interface; due to density and viscosity differences between the fluids, the interface may take on a curved shape where the injected and resident fluids are vertically segregated.  These interface geometries result in $r_{o}$ values which are a function of $z_{o}$ and time in general. Figure \ref{fig:one} depicts the motion of an arbitrarily shaped immiscible interface over a very short time interval $\Delta t$, travelling with radial velocity $v=A/r$, where $A$ is constant with respect to $r$. Note that the dummy coordinates $r_{o},z_{o}$ are not used in the figure as it is depicting the time dependent, “true” position of the interface rather than the effective intersection surface which intersects an arbitrary point. If the overall shape of the interface does not significantly change over this interval, then the interface’s height at $t+\Delta t$ is approximately the same as the height of the interface at time $t$ but at a postion further back by $\Delta r = v\Delta t$, i.e.:

\begin{figure} 
  \centerline{\includegraphics[scale=0.7]{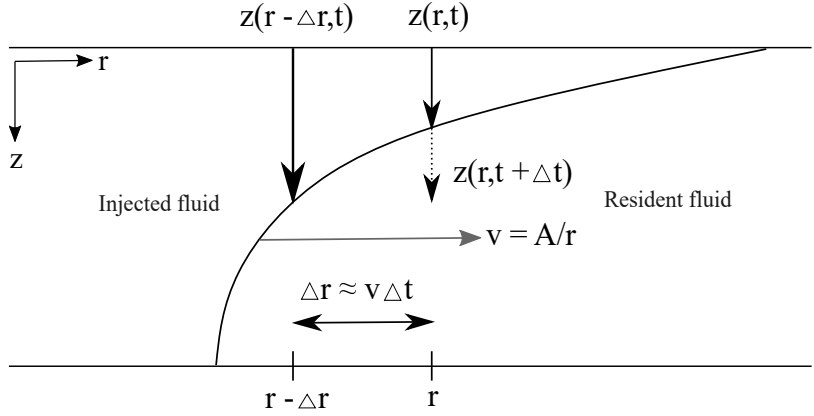}}
  \caption{An immiscible interface travelling over a short time interval $\Delta t$}
\label{fig:one}
\end{figure}

\begin{equation}
  z(r,t+\Delta t) \approx z(r-\Delta r,t)
\end{equation}

This relationship can be used in approximating the time derivative of the interface height $z$:

\begin{equation}
  \frac{\partial z}{\partial t} \approx \frac{z(r-\Delta r,t)-z(r,t)}{\Delta t} = v\left(\frac{z(r-\Delta r,t)-z(r,t)}{\Delta r}\right) \approx -v\frac{\partial z}{\partial r}
\end{equation}

Thus,

\begin{equation}
  \frac{\partial z}{\partial t} + v\frac{\partial z}{\partial r} = 0
\label{dzEqn}
\end{equation}

By adding and subtracting a term $zv/r$ to equation \ref{dzEqn}, one can obtain a continuity equation for the height of the immiscible interface with respect to distance and time:

\begin{equation}
  \frac{\partial z}{\partial t} + \frac{1}{r}\frac{\partial}{\partial r}(rzv) = 0
\label{eqnTwoNine}
\end{equation}

The resident and injected fluids are both assumed to have hydrostatic fluid pressure and velocity fields governed by Darcy's law (i.e. proportional to the pressure gradient of each fluid). The mathematical expressions for these assumptions can be combined with equation \ref{eqnTwoNine} to obtain the governing differential equation for the interface's evolution:

\begin{equation} 
  \frac{\partial z}{\partial t} - \frac{1}{r\phi}\left(\frac{g\Delta\rho kz(H-z)r}{\mu_{r}z + \mu_{i}(H-z)}\frac{\partial z}{\partial r} + \frac{Q\mu_{i}(H-z)}{2\pi(\mu_{r}z + \mu_{i}(H-z))}\right) = 0
\end{equation}

The $\mu_{r}$ and $\mu_{i}$ terms are the viscosities of the residual and injected fluids respectively, $\Delta\rho$ is their density difference, $H$ is the height of the aquifer, $k$ is its permeability, and $g$ is the acceleration due to gravity. This form of the equation appears in \citet{NordbottenCelia06} as well as \citet{Guoetal16}, who derive approximate solutions for it given various fluid properties.

Given an analytical expression for a solution for this governing equation (typically $z$ as a function of $r,t$) and solving for the intersection time $t_{o}$ at a point $(r,z)$, one can apply the effective intersection surface integral in equation \ref{Qflux} by substituting $t_{o}$ back into the solution in terms of the dummy coordinates $r_{o}$ and $z_{o}$, and solving for $r_{o}$ in terms of $r,z,$ and $z_{o}$. Finally, this expression can be used in the flux integral to obtain an expression for the velocity field, which will be of the incompressible form:

\begin{equation}
  \boldsymbol{v} = \frac{A(z)}{r}\hat{\boldsymbol{r}}
\end{equation}

In general, the term $A$ is a function of $z$ characteristic of the analytical interface solution. In the case of the cylindrical interface, $A$ is a constant equal to $Q/2\pi\phi H$. 

In reality, the interface between injected and residual fluids is never sharp. There is always dispersion of the fluids due to mass transfer effects taking place at the interface. Such processes are governed by the advection-diffusion equation (ADE), which has the general form \citep{Stocker11}: 

\begin{equation}
  \frac{\partial c}{\partial t} = \nabla\cdot(\boldsymbol{D}\nabla c) - \nabla\cdot(\boldsymbol{v}c) + P
\end{equation}

Here, $c$ is the fluid concentration, $\boldsymbol{D}$ is the diffusivity (a tensor, in general; see \citet{Bear61}) and $P$ represents sources and sinks. In the domain of the aquifer outside of the injection site, it is assumed that $P=0$. The diffusivity term is taken to be the sum of the molecular diffusion and mechanical dispersion effects, as done by \citet{Neumanetal87}:

\begin{equation} 
  D_{T} = D_{m}+d_{T}\lVert\boldsymbol{v}\rVert, \quad D_{L} = D_{m}+d_{L}\lVert\boldsymbol{v}\rVert
\end{equation}

Here, $D_{m}$ is the molecular diffusion coefficient (assumed constant), $D_{T}$ and $D_{L}$ are the transverse (normal to the velocity field) and longitudinal (parallel to the velocity field) diffusivity components respectively, and $d_{T}$ and $d_{L}$ are the transverse and longitudinal dispersivity scales. In the present study's case of a strictly radial velocity field in an isotropic medium, the transverse dispersivity is taken to be 0, and the longitudinal dispersivity $d_{L}$ is simply referred to as the dispersivty $d$. It has been suggested that in large field scale transport situations, longitudinal dispersivity approaches a constant or asymptotic value at larger distances \citep{PickensGrisak81}. The asymptotic value (which is constant) can serve as a worst case scenario for evaluation of dispersion. Furthermore, if the effects of molecular diffusion are assumed to be far less than those of mechanical dispersion (a very good assumption for all practical applications), the ADE can be simplified as: 

\begin{equation}
  \frac{\partial c}{\partial t} + \frac{A}{r}\left(\frac{\partial c}{\partial r}-d\frac{\partial^{2}c}{\partial^{2}r}\right) = 0
\label{ADELite}
\end{equation}

This form of the ADE and its solutions have been the subject of numerous hydrology studies. One of a few approximate solutions, given by \citet{Dagan71} and \citet{Hsieh86} is of practical interest as it has a tendency to slightly exceed numerical solutions to equation \ref{ADELite}, thus acting as an ``upper bound'' estimate on the concentration profile in the aquifer. This solution has the dimensional form:

\begin{equation}
  c(r,t) = \frac{c_{o}}{2}\mathrm{erfc}\left\{\frac{\left(\ln\left(\frac{r}{d}\right) - \ln\left(\frac{1}{d}\sqrt{2At+r_{o}^{2}}\right)\right)\left(2At+r_{o}^{2}\right)}{d^{2}\sqrt{\frac{4}{3d^{3}}\left((2At+r_{o}^{2})^{\frac{3}{2}}-r_{o}^{3}\right)}}\right\}
\end{equation}

In this equation, $r_{o}$ is the radius of the injection well, $c_{o}$ is the initial concentration of the injected fluid, and $A$ is the term from the incompressible velocity field. For an aquifer of infinite radial extent, the injection well radius $r_{o}$ can be neglected, giving the so called ``line solution'' as done by \citet{Guoetal16}. The above equation can also be solved for $r$ as a function of time and concentration, which is of practical interest and allows for comparing the “cut off” for the extent of injected fluid presence with an immiscible interface. After finding the velocity field for general interface geometries, the $A(z)$ term can be substituted into this equation to allow for $z$ dependent solutions: this procedure is carried out below for some analytical interface solutions.

\section{Results}
\label{sec:res}

To demonstrate this transformation, three approximate immiscible interface solutions for $z$ as a function of $r,t$ from \citet{Guoetal16} have been chosen for its application. For all these solutions, the injected fluid is assumed to be less dense than the resident fluid, while the driving effects of buoyancy are assumed to be far less than those of injection, with viscosity differences between the injected and resident fluids being the key factor in each solution’s geometry. All three solutions are fixed to be $z=0$ and $z=H$ outside of two moving boundaries characteristic of each solution. For the intersection surface used in the transformation, only the solution in the region within these boundaries is used as it represents the travelling interface.

\subsection{Injected Fluid is More Viscous than Resident Fluid}

Under these circumstances, \citet{Guoetal16} provides the following “travelling interface” height solution:

\begin{equation}
  z(r,t) = \frac{H(M-1)}{2M\Gamma}\left(\frac{\pi\phi Hr^{2}}{Qt}-1\right)+\frac{H}{2}, \qquad 1-\frac{M}{1-M}\Gamma < \frac{\pi\phi Hr^{2}}{Qt} \leq 1+\frac{M}{1-M}\Gamma
\label{MLT1}
\end{equation}

Here, the dimensionless parameters $\Gamma = 2\pi\Delta\rho gkH^{2}/\mu_{r}Q$ represents the effect of buoyancy compared to injection and $M=\mu_{r}/\mu_{i}$ is the ratio of the resident and injected fluids' viscosities. Mathematically, the solution in \ref{MLT1} assumes that $\Gamma \ll 1$ and $M<1$.

Let $t=t_{o}$ be the time a specific point $(r,z)$ is intersected by the travelling interface. Rearranging \ref{MLT1} for this time yields:

\begin{equation}
  t_{o} = \frac{\pi\phi Hr^{2}}{Q\left(1+\frac{2M\Gamma}{M-1}\left(\frac{z}{H}-\frac{1}{2}\right)\right)}
\end{equation}

Substituting this intersection time back into the solution with respect to the dummy variables:

\begin{equation}
  z_{o} = \frac{H(M-1)}{2M\Gamma}\left(\frac{r_{o}^{2}}{r^{2}}\left(1+\frac{2M\Gamma}{M-1}\left(\frac{z}{H}-\frac{1}{2}\right)\right) - 1\right)+\frac{H}{2}
\end{equation}

Solving for $r_{o}$ yields:

\begin{equation}
  r_{o} = r\sqrt{\frac{1+\frac{2M\Gamma}{M-1}\left(\frac{z_{o}}{H}-\frac{1}{2}\right)}{1+\frac{2M\Gamma}{M-1}\left(\frac{z}{H}-\frac{1}{2}\right)}}
\end{equation}

Substituting this dummy variable $r_{o}$ as a function of $r,z$ and $z_{o}$ into \ref{vDA} and then into the flux integral (\ref{Qflux}):

\begin{equation}
  Q = \frac{2\pi\phi vr}{\sqrt{1+\frac{2M\Gamma}{M-1}\left(\frac{z}{H}-\frac{1}{2}\right)}}\int_{0}^{H}\sqrt{1+\frac{2M\Gamma}{M-1}\left(\frac{z_{o}}{H}-\frac{1}{2}\right)}dz_{o}
\end{equation}

Finally, evaluating this integral and solving for the velocity field yields:

\begin{equation}
  v(r,z) = \frac{3QM\Gamma\sqrt{1+\frac{2M\Gamma}{M-1}\left(\frac{z}{H}-\frac{1}{2}\right)}}{2\pi\phi rH(M-1)\left(\left(1+\frac{M\Gamma}{M-1}\right)^{\frac{3}{2}}-\left(1-\frac{M\Gamma}{M-1}\right)^{\frac{3}{2}}\right)}
\end{equation}

As expected, this velocity field is of the incompressible form:

\begin{equation}
  \boldsymbol{v} = \frac{A(z)}{r}\hat{\boldsymbol{r}}, \qquad A(z) = \frac{3QM\Gamma\sqrt{1+\frac{2M\Gamma}{M-1}\left(\frac{z}{H}-\frac{1}{2}\right)}}{2\pi\phi H(M-1)\left(\left(1+\frac{M\Gamma}{M-1}\right)^{\frac{3}{2}}-\left(1-\frac{M\Gamma}{M-1}\right)^{\frac{3}{2}}\right)}
\label{AforMLT1}
\end{equation}

Using this velocity field, the 2-D concentration profile can be given by substituting $A(z)$ above into the equation from \citet{Dagan71}. Similarly, one can also rearrange the concentration equation for $r$, and find the radial extent of boundaries that are a function of a specific concentration. Two boundaries of practical interest are where $c/c_{o} = 0.99$, and $c/c_{o} = 0.01$, as these provide an approximation for a “transition zone” where the relative concentration of the injected fluid experiences the most variation due to dispersion; outside of these boundaries, the aquifer is essentially saturated with the resident or the injected fluid. The inverted concentration equation (including the line solution approximation $r_{o} \approx 0$) has the form:

\begin{equation}
  r(c,z,t) = \sqrt{2A(z)t}\mathrm{exp}\left\{ \frac{2}{(2A(z)t)^{\frac{1}{4}}}\sqrt{\frac{d}{3}} \mathrm{erfc}^{-1}\left\{\frac{2c}{c_{o}}\right\} \right\}
\label{RofCZT}
\end{equation}

To demonstrate the effect of mechanical dispersion, consider the radial extent of the $c/c_{o} = 0.01$ boundary at the top of the aquifer (at $z=0$ for a positive-downward z-axis) where the immiscible interface solution in \ref{MLT1} is at its furthest radial extent; let these two radial extents be denoted $r_{1}$ and $r_{2}$ respectively. Furthermore, let the function $f(M,\Gamma)$ be the collected ``$M$ terms'' from $A(0)$ in \ref{AforMLT1}, i.e.:

\begin{equation}
  A(0) = \frac{3Q\Gamma}{2\pi\phi H}f(M,\Gamma), \qquad f(M,\Gamma) = \frac{M\sqrt{1-\frac{M\Gamma}{M-1}}}{(M-1)\left(\left(1+\frac{M\Gamma}{M-1}\right)^{\frac{3}{2}}-\left(1-\frac{M\Gamma}{M-1}\right)^{\frac{3}{2}}\right)}
\end{equation}

This allows the term $r_{1}$ to be expressed as:

\begin{equation}
  r_{1} = \sqrt{\frac{3Q\Gamma tf(M,\Gamma)}{\pi\phi H}}\mathrm{exp}\left\{ \frac{2}{\left(\frac{3Q\Gamma tf(M,\Gamma)}{\pi\phi H}\right)^{\frac{1}{4}}}\sqrt{\frac{d}{3}} \mathrm{erfc}^{-1}\left\{0.02\right\} \right\}
\label{MLT1:r1Dimd}
\end{equation}

Furthermore, let the dimensionless parameters for time and radial distance be $T$ and $R$ respectively, defined as:

\begin{equation}
  T = \frac{Qt}{d^{2}\pi\phi H}, \qquad R = \frac{r}{d}
\end{equation}

Using these parameters, equation \ref{MLT1:r1Dimd} can be nondimensionalized as:

\begin{equation}
  R_{1} = \sqrt{3T\Gamma f(M,\Gamma)}\mathrm{exp}\left\{ \frac{2\mathrm{erfc}^{-1}\left\{0.02\right\}}{\sqrt{3}}(3T\Gamma f(M,\Gamma))^{\frac{-1}{4}} \right\}
\end{equation}

The term $r_{2}$ can be found by solving the upper bound on the domain in \ref{MLT1} for $r$:

\begin{equation}
  r_{2} = \sqrt{\frac{Qt}{\pi\phi H}\left(1+\frac{M\Gamma}{1-M}\right)}
\end{equation}

Which can be similarly nondimensionalized to obtain the expression:

\begin{equation}
  R_{2} = \sqrt{T\left(1+\frac{M\Gamma}{1-M}\right)}
\end{equation}

Finally, the dimensionless difference between the 1\% concentration boundary and the $M<1$ immiscible solution positions at $z=0$ can be written as:

\begin{equation}
  R_{1} - R_{2} = \sqrt{T}\left(\sqrt{3\Gamma f(M,\Gamma)}\mathrm{exp}\left\{ \frac{2\mathrm{erfc}^{-1}\left\{0.02\right\}}{\sqrt{3}}(3T\Gamma f(M,\Gamma))^{\frac{-1}{4}} \right\} - \sqrt{\left(1+\frac{M\Gamma}{1-M}\right)}\right)
\label{MLT1Diff}
\end{equation}

Note that the function $R_{1}-R_{2}$ in \ref{MLT1Diff} is not well defined on the whole interval $M\epsilon(0,1)$. The fucntion $f(M,\Gamma) = 0$ when $M=1/(1+\Gamma)$, which causes the exponential term in $R_{1}-R_{2}$ to diverge. Thus, the domain of \ref{MLT1Diff} is $M\epsilon(0,1/(1+\Gamma))$, the upper boundary of which approaches 1 in the limit of $\Gamma \rightarrow 0$.

\begin{figure}
  \centerline{\includegraphics[scale=0.9]{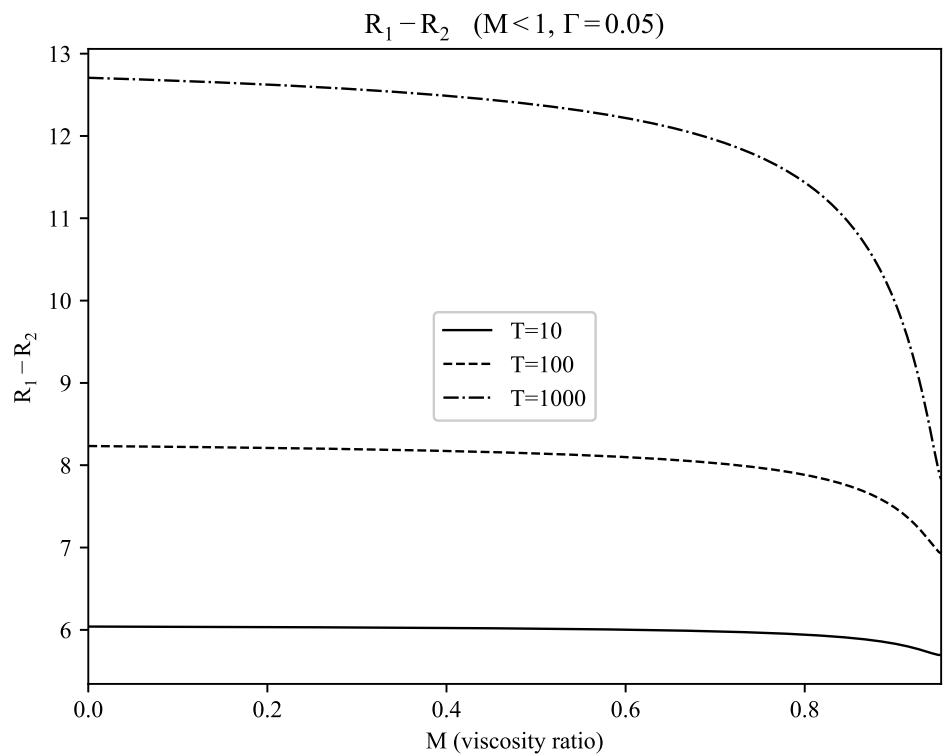}}
  \caption{The dimensionless difference between the 1\% relative concentration boundary and the immiscible interface solution at the top of the aquifer $(M<1)$}
\label{fig:two}
\end{figure}

The plot in figure \ref{fig:two} shows this dimensionless separation between the 1\% boundary and the immiscible interface for $\Gamma = 0.05$ and $T=10,100$ and $1000$. The dimensionless separation function is always positive, and though its time derivative is initially negative, the separation increases with time after $T\approx2$; this increased separation with time is illustrated in the plot. 

\subsection{Injected Fluid is Less Viscous than Resident Fluid}

The approximate solution provided by \citet{Guoetal16} for a more viscous resident fluid $(M>1)$ is:

\begin{equation}
  z(r,t) = \frac{H}{M-1}\sqrt{\frac{QMt}{\pi\phi Hr^{2}}}-H, \qquad \frac{Q}{\pi\phi HM} < \frac{r^{2}}{t} \leq \frac{QM}{\pi\phi H}
\label{MGT1}
\end{equation}

This produces the $r_{o}$ expression:

\begin{equation}
  r_{o}(r,z,z_{o}) = r\left(\frac{1+\frac{z(M-1)}{H}}{1+\frac{z_{o}(M-1)}{H}}\right)
\end{equation}

Evaluating the flux integral with this $r_{o}$ expression yields the velocity field:

\begin{equation}
  \boldsymbol{v} = \frac{A(z)}{r}\hat{\boldsymbol{r}}, \qquad A(z) = \frac{Q(M-1)}{2\pi\phi(z(M-1)+H)\ln{M}}
\label{318}
\end{equation}

Once again, consider the radial extents (at the top of the aquifer) of the 1\% relative concentration boundary ($r_{1}$) and the immiscible solution ($r_{2}$) for the $M>1$ case. Evaluating \ref{318} at $z=0$ with equation \ref{RofCZT} gives the following expression for $r_{1}$:

\begin{equation}
  r_{1} = d\sqrt{\frac{Qt(M-1)}{\pi\phi H\ln{M}}}\mathrm{exp}\left\{ \frac{2\mathrm{erfc}^{-1}(0.02)}{\sqrt{3}}\left(\frac{Qt(M-1)}{\pi\phi H\ln{M}}\right)^{-\frac{1}{4}} \right\}
\end{equation}

The upper boundary on the domain of the solution \ref{MGT1} can be solved again to find $r_{2}$:

\begin{equation}
  r_{2} = d\sqrt{\frac{QMt}{\pi\phi H}}
\end{equation}

Nondimensionalizing these terms with the parameters $T$ and $R$ as defined above gives the dimensionless separation for the 1\% boundary and the immiscible solution at $z=0$ for the $M>1$ case:

\begin{equation}
  R_{1} - R_{2} = \sqrt{T}\left(\sqrt{\frac{M-1}{\ln{M}}}\mathrm{exp}\left\{ \frac{2\mathrm{erfc}^{-1}(0.02)}{\sqrt{3}}\left(\frac{T(M-1)}{\ln{M}}\right)^{-\frac{1}{4}} \right\} - \sqrt{M} \right)
\label{MGT1Diff}
\end{equation}

This separation function only gives physical solutions on a restricted domain. For certain $M$ and $T$ values, equation \ref{MGT1Diff} will be $\leq 0$, which would correspond to a concentration boundary lagging behind the underlying immiscible solution. Additionally, the derivative of \ref{MGT1Diff} with respect to $T$  has a root of its own at a time dependent $M$ value between 1 and the original function's root. While $M$ values in between these roots correspond to a 1\% boundary that lies ahead of the original interface, this case results in the interface converging toward and then exceeding the concentration boundary. As such, realistic solutions for the $M>1$ case are restricted between $M=1$ and the root of the separation's time derivative, as the 1\% boundary will recede from the interface in this range. 

\begin{figure} 
  \centerline{\includegraphics[scale=0.9]{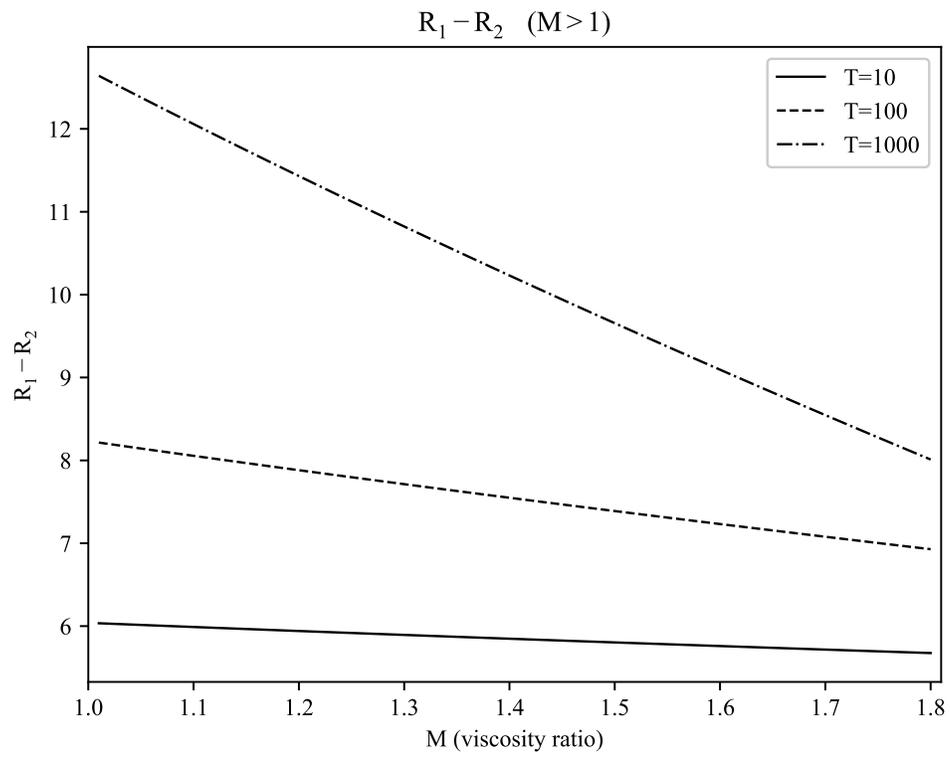}}
  \caption{The dimensionless difference between the 1\% relative concentration boundary and the immiscible interface solution at the top of the aquifer $(M>1)$}
\label{fig:thr}
\end{figure}

Over time, the derivative's root will approach $M=1$, thus solutions that occur before earlier $T$ values have larger $M$ domains. The $M$ value of this root at a certain $T$ can be found numerically. For example, cases that occur up to $T=10$ have physical solutions approximately on $M\epsilon(1,3.87)$, while cases up to $T=1000$ are restricted to $M\epsilon(1,1.81)$; the latter domain is used to plot the dimensionless separation of the 1\% boundary and the original interface over time in figure \ref{fig:thr}.

\subsection{Equal Viscosity for Injected Fluid and Resident Fluid} \label{3p3}

As a final example, consider the case for fluids of equal viscosity. The approximate solution provided by \citet{Guoetal16} is:

\begin{equation}
  z(r,t) = \frac{H}{2}\left(1-\frac{1}{\sqrt{\Gamma}}\left(\frac{\pi\phi Hr^{2}}{Qt}-1\right)\right), \qquad \frac{Q}{\pi\phi H}(1-\sqrt{\Gamma}) < \frac{r^{2}}{t} \leq \frac{Q}{\pi\phi H}(1+\sqrt{\Gamma})
\end{equation}

As before, solving for the intersection time $t_{o}$ to get the dummy variable $r_{o}$ yields: 

\begin{equation}
  r_{o}(r,z,z_{o}) = r\sqrt{\frac{2+4\sqrt{\Gamma}\left(\frac{1}{2}-\frac{z_{o}}{H}\right)}{2+4\sqrt{\Gamma}\left(\frac{1}{2}-\frac{z}{H}\right)}}
\end{equation}

Carrying out the same flux integral as before provides the velocity field:

\begin{equation}
  \boldsymbol{v} = \frac{A(z)}{r}\hat{\boldsymbol{r}}, \qquad A(z) = \frac{3Q\sqrt{\Gamma\left(2+4\sqrt{\Gamma}\left(\frac{1}{2}-\frac{z}{H}\right)\right)}}{\pi\phi H\left(\left(2+2\sqrt{\Gamma}\right)^{\frac{3}{2}}-\left(2-2\sqrt{\Gamma}\right)^{\frac{3}{2}}\right)}
\end{equation}

Evaluating the function $A(z)$ at $z=0$ can be written in terms of a function $g(\Gamma)$, written as:

\begin{equation}
  A(0) = \frac{3Q}{\pi\phi H}g(\Gamma), \qquad g(\Gamma) = \frac{\sqrt{\Gamma(2+2\sqrt{\Gamma})}}{(2+2\sqrt{\Gamma})^{\frac{3}{2}} - (2-2\sqrt{\Gamma})^{\frac{3}{2}}}
\end{equation}

Finally, defining the dimensionless quantites quantities $R_{1}$ and $R_{2}$ the same way as before, the $M=1$ case has the following dimensionless separation:

\begin{equation}
  R_{1} - R_{2} = \sqrt{T}\left(\sqrt{6g(\Gamma)}\mathrm{exp}\left\{ \frac{2\mathrm{erfc}^{-1}(0.02)}{\sqrt{3}\left(6Tg(\Gamma)\right)^{\frac{1}{4}}}\right\} - \sqrt{1+\sqrt{\Gamma}} \right)
\end{equation}

\begin{figure}
  \centerline{\includegraphics[scale=0.9]{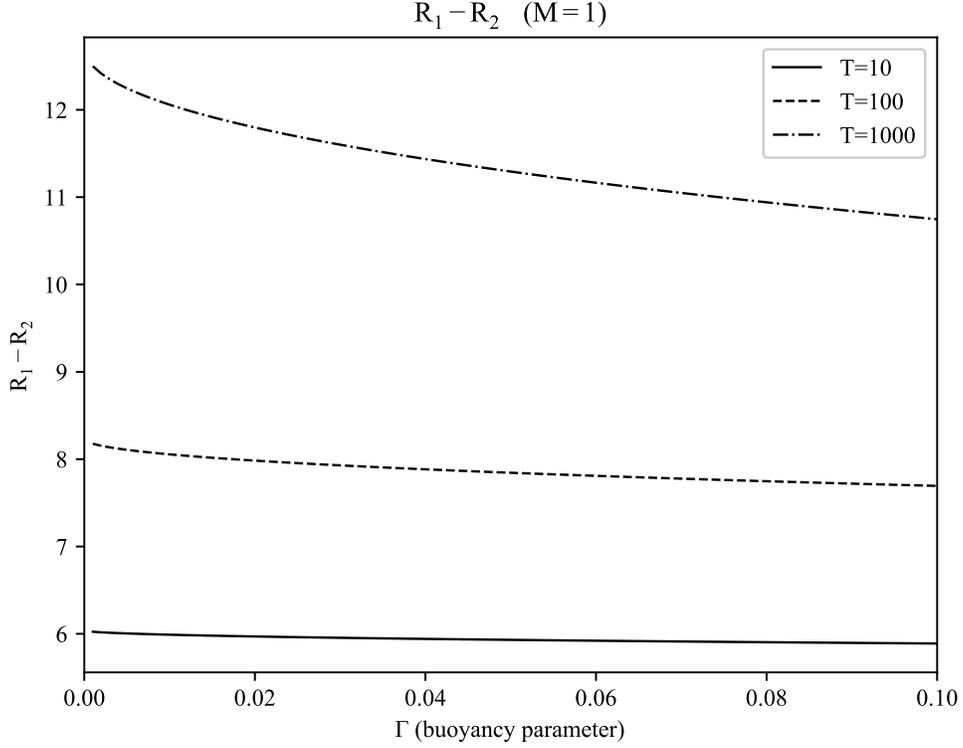}}
  \caption{The dimensionless difference between the 1\% relative concentration boundary and the immiscible interface solution at the top of the aquifer $(M=1)$}
\label{fig:four}
\end{figure}

This separation function is well defined for $\Gamma\epsilon(0,1)$ and is always greater than zero. Like the $M<1$ case, the time derivative of this function is initially negative but the separation proceeds to increase monotonically with respect to time after $T \approx 2$ for $\Gamma \ll 1$ (this transition time varies slightly with $\Gamma$, being $T\approx2.009$ for $\Gamma=0.01$, and $T\approx1.994$ for $\Gamma=0.1$). The dimensionless distance between the 1\% relative concentration boundary and the immiscible interface solution is plotted in figure \ref{fig:four}.  

\begin{figure} 
  \centerline{\includegraphics[scale=0.9]{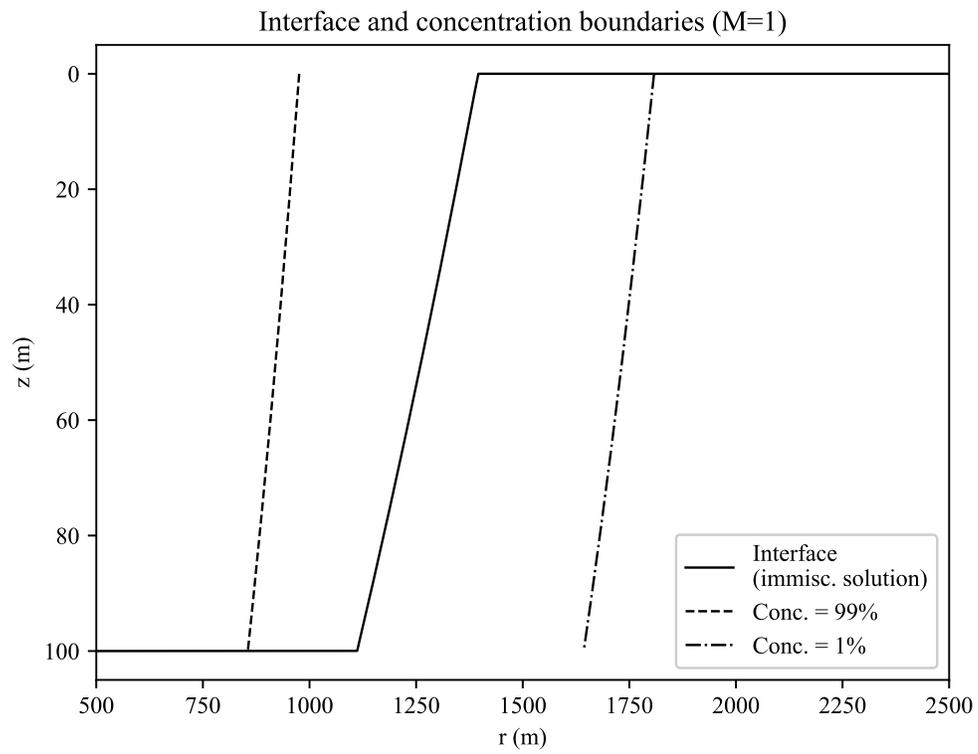}}
  \caption{The solution for the $M=1$ immiscible interface, with the boundaries for 1\% and 99\% relative concentration of the injected fluid }
\label{fig:five}
\end{figure}

As a final demonstration of the concentration boundaries produced by mechanical dispersion, consider the following example with physical values. The $M=1$ case can be used to represent injecting aqueous carbon dioxide into an aquifer for CCS. One of the motivations for the present paper is developing a novel CCS enhancement which allows for significantly reducing or eliminating the risks of CO$_{2}$ leakage. In this aproach, instead of injecting pure CO$_{2}$, it is dissolved in brine produced from a target aquifer and reinjected back into the formation \citep{Caoetal20, Caoetal21}. This problem's formulation becomes identical to injecting contaminant fluid in porous media, but the importance of despersion is increased due to much larger scales of injection. 
Typical aquifer parameters and a typical CCS project injection rate and duration are given in table \ref{ParamTable}, and figure \ref{fig:five} illustrates an example of these transition zone boundaries ($c/c_{o} = 0.99$ and $c/c_{o} = 0.01$) compared to the $M=1$ immiscible interface given by \citet{Guoetal16}. 

\begin{table}
  \begin{center}
\def~{\hphantom{0}}
  \begin{tabular}{llr}
       Variable  & Quantity [Dimensions]  & Value\\[3pt]
       $Q$   & Volume injection rate [L$^{3}$T$^{-1}$] & 1,000,000 m$^{3}$/year\\
       $H$   & Aquifer height [L] & 100 m\\
       $M$   & Viscosity ratio (res./inj. fluid) & 1\\
       $\Gamma$   & Buoyancy parameter & 0.05\\
       $\phi$ & Aquifer Porosity & 0.1\\
       $t$ & Time elapsed [T] & 50 years\\
       $d$ & Dispersivity scale [L] & 35 m\\
  \end{tabular}
  \caption{List of CCS quantities used to plot physical interface and transition boundaries.}
  \label{ParamTable}
  \end{center}
\end{table}

It is clear from the plot in figure \ref{fig:five} that the concentration transition zone due to mechanical dispersion is considerably spread around the underlying immiscible interface solution, with the 1\% and 99\% boundaries separated by several hundred metres. As in \citet{Guoetal16}, the injected fluid is assumed to be less dense than the groundwater, and the z-axis is positive downward. The radial extent of the interface and the boundaries (to the nearest metre) at the top and bottom of the aquifer are summarized in table \ref{TabTwo}:

\begin{table}
  \begin{center}
\def~{\hphantom{0}}
  \begin{tabular}{llll}
       & 99\% boundary & Interface solution & 1\% boundary\\[3pt]
       At $z=0$ m (top): & $r=976$ m & $r=1395$ m & $r=1808$ m\\
       At $z=100$ m (bottom): & $r=855$ m & $r=1112$ m & $r=1643$ m\\
  \end{tabular}
  \caption{Example radial extents for the $M=1$ Interface, 1\% and 99\% concentration boundaries at the top and bottom of the aquifer.}
  \label{TabTwo}
  \end{center}
\end{table}

\section{Discussion}

It is clear that mechanical dispersion can allow for an injected fluid concentration transition zone that is spread considerably on either side of an immiscible interface solution. For the three examples used throughout section \ref{sec:res}, the 1\% relative concentration boundary location was found to increasingly recede from the original interface solution over time, subject to domain restrictions for physical solutions. This is consistent with the interpretation of immiscible interfaces acting as effective surfaces underlying a dispersive concentration profile. 

The three immiscible interface cases examined were solutions from \citet{Guoetal16}, where the viscosity differences between the injected fluid and resident fluid were the primary factor determining the interfaces’ geometries. For all three cases, the dimensionless difference between the 1\% boundary and the original interface at the top of the aquifer $R_{1} - R_{2}$ is found to lie approximately in the range 6-13, with larger differences occuring at later dimensionless times $T$. This corresponds to 1\% concentration traces of the injected fluid for a given interface geometry lying 6-13 dispersivity lengths ahead of the immiscible interface solution. This observation is also consistent with the “effective surface” interpretation of the concentration profile, as the 1\% relative concentration boundary of injected fluid due to mechanical dispersion is found to propagate further than the furthest radial extent of the underlying immiscible interface solution (which occurs at $z=0$).

For the $M<1$ and $M=1$ cases, the dimensionless separation between the 1\% boundary and the interface $R_{1} - R_{2}$ is strictly positive but initially has a negative time derivative, which becomes positive for $T\gtrsim2$. Physically, this corresponds to the original interface's position initially approaching closer to the 1\% concentration boundary (but never reaching it), before the boundary begins to recede away from the interface position, with increasing distance between them over time.

Subject to more complicated domain ($M$ value) restrictions, the $M>1$ case also exhibits this behaviour. However, the $M>1$ case in particular highlights the limits of the ``effective surface'' interpretation's applicability. There are time dependent values of the viscosity ratio for the $M>1$ case for which the 1\% concentration boundary would ``lag behind'' the original interface, and for these cases it would be better to consider the immiscible solution as the ``worst case scenario'' for the extent of the injected fluid's presence. For cases of sufficiently small injection duration/$T$ values and large $M$ values (or vice versa), the effective surface interpretation should be applied and the 1\% relative concentration boundary should be considered the furthest extent of the injected fluid. 

Thus the results in section \ref{sec:res} illustrate the need to account for mechanical dispersion in the fluids' evolution in the aquifer for many practical cases. Note that the relative concentration $c/c_{o}$ itself is a variable quantity which affects the applicability of considering a concentration boundary the worst case scenario for fluid presence. A relative concentration of 1\% was considered and treated as a constant value throughout section \ref{sec:res} , but is by no means unique. For the $M>1$ case, considering smaller relative concentrations (< 1\%) to evaluate the position of a boundary expands the applicability of the procedure derived above (i.e. the concentration boundary will recede from the immiscible solution for larger $M$ values). Likewise, considering relative concentrations larger than 1\% diminishes the range of applicability for the ``effective surface'' concentration profile. For all three viscosity cases treated above, it can be shown that concentration boundaries of a sufficiently large relative concentration in fact lie behind the corresponding immisicible solution, rather than receding away from it. In any case, physical restrictions are a necessary consequence of considering simplified, analytical solutions.  

The location of the 1\% relative concentration boundary considered in section \ref{sec:res} is particularly useful for injection well engineering applications. For example, if one wanted to know the extent of an injected contaminant’s presence after a period of time, the immiscible interface solution would tend to underestimate the upper limit of its radial position; as seen in the plots, the immiscible solution lies within the concentration transition zone due to mechanical dispersion. In practice, if aquifer water were to be sampled from just outside the furthest extent of the immiscible solution at a given height in the aquifer, it could still contain considerable traces (> 1\%) of the injected contaminant. For safety measures, the 1\% boundary location (or the boundary of a desired cutoff concentration, e.g. 5\%, 0.1\%, etc., which can be determined using the same methodology that has been outlined) should be treated as a worst-case scenario for sampling uncontaminated water. 

A physical example is given in section \ref{3p3} for the equal viscosity case, which can be applied to an aqueous carbon dioxide solution being injected for CCS. From the results summarized in table \ref{TabTwo}, the 99\% boundary is found to lie 257-419 m behind the immiscible solution, and the 1\% boundary is found to lie 413-531 m ahead of the immiscible solution using the example values in table \ref{ParamTable}. Comparing these physical results to the nondimensional ones for $M=1$, the difference between the interface and the 1\% boundary at $z=0$ (413 m) is 11.8 times the dispersivity scale $d = 35$m after 50 years of injection (corresponding to a dimensionless duration of $T\approx1300$); this is consistent with the 6-13 range for the dimensionless separation for a similar duration seen in figure \ref{fig:four}.

\section{Conclusions}

The goal of this study was to produce a mathematical technique for accounting the effect of mechanical dispersion on the evolution of injected fluid interfaces of an arbitrary geometry (non-cylindrical) travelling through a confined, porous aquifer. Given an analytic expression for an immiscible solution, the interface will intersect a specific point in the aquifer $(r,z)$ at a time $t_{o}$ with a specific shape given by the solution. If the set of intersecting interfaces throughout the aquifer’s domain is treated as a set of effective surfaces to evaluate the volumetric flux of the injected fluid through, an incompressible, time independent velocity field $\boldsymbol{v} = \frac{A(z)}{r}\hat{\boldsymbol{r}}$ can be obtained, where the function $A(z)$ is characteristic of the underlying immiscible solution. This velocity field is then used to solve the ADE and obtain the concentration profile of the injected fluid throughout the aquifer, which shows that significant amounts of injected fluid can be present beyond the extent of the original interface solution. 

One result of this technique of practical interest is to obtain the location of a boundary of a desired relative concentration. For real world applications involving contaminant injection into aquifers, such boundaries allow for defining a “cut off” position of arbitrarily low relative concentration, beyond which the groundwater may be considered uncontaminated. This will allow for engineers to plan injection well sites more cautiously for these applications and quantify “worst case scenarios” for contaminant presence away from the injection well.

It is worth mentioning that the choice of using the equation from \citet{Dagan71} and its inversion for the concentration and radial extent of the cutoff boundaries is not unique. Approximate solutions for the ADE exist in other forms, such as that given by \citet{TangBabu79}. Additionally, the ADE with the velocity field described above and negligible molecular diffusion is straightforward to solve by numeric means and doing so can serve the same purposes outlined above. The general nature of the transformation outlined in this study should in principle allow for any immiscible interface and ADE solutions to be used to obtain a dispersive concentration profile throughout an aquifer.

\end{document}